\RequirePackage{lineno}
\documentclass[prb,aps,twocolumn, showpacs,amssymb]{revtex4}
\usepackage{t1enc}
\usepackage[latin1]{inputenc}
\usepackage[english]{babel}
\usepackage{graphics}
\usepackage{graphicx}
\usepackage{amssymb}
\usepackage{amsmath}
\usepackage{dcolumn}
\usepackage{bm}
\usepackage{color}
\usepackage{verbatim}  
\usepackage{vector}  
\usepackage{wrapfig}
\usepackage{enumerate}
\usepackage{sidecap}

\usepackage{lineno}

\begin{document}

\renewcommand{\figurename}{Fig.}

\title{A weak-coupling superconductivity in electron doped NaFe$_{0.95}$Co$_{0.05}$As is revealed by ARPES}
\author{S.\ Thirupathaiah,$^1$ D. V.\ Evtushinsky,$^1$  J.\ Maletz, $^1$ V. B.\ Zabolotnyy,$^1$  A. A.\ Kordyuk,$^{1,3}$  T. K.\ Kim,$^2$ S. Wurmehl,$^1$ M. Roslova,$^{1,4}$ I. Morozov,$^{1,4}$ B.\ {B{\"u}chner},$^{1,5}$  and S. V.\ Borisenko,$^1$}

\affiliation{
$^1$Leibniz-Institute for Solid State and Materials Research Dresden, P.O.Box 270116, D-01171 Dresden, Germany\\
$^2$Diamond Light Source Ltd., Didcot, Oxfordshire, OX11 0DE, United Kingdom\\
$^3$Institute of Metal Physics of National Academy of Sciences of Ukraine, 03142 Kyiv, Ukraine\\
$^4$Department of Chemistry, Moscow State University, 119991 Moscow, Russia\\
$^5$Institut f$\ddot{u}$r Festk$\ddot{o}$rperphysik, Technische Universit$\ddot{a}$t Dresden, D-01171 Dresden, Germany}
\date{\today}

\begin{abstract}
We report a systematic study on the electronic structure and superconducting (SC) gaps in electron doped NaFe$_{0.95}$Co$_{0.05}$As superconductor using angle-resolved photoemission spectroscopy. Hole-like Fermi sheets are at the zone center and electron-like Fermi sheets are at the zone corner, and are mainly contributed by $xz$ and $yz$ orbital characters. Our results reveal a $\frac{\Delta}{K_B T_c}$ in the range of 1.8-2.1, suggesting a weak-coupling superconductivity in these compounds.  Gap closing above the transition temperature ($T_c$) shows the absence of pseudogaps. Gap evolution with temperature follow the BCS gap equation near the $\Gamma$, $Z$, and $M$ high symmetry points. Furthermore, an almost isotropic superconductivity along $k_z$ direction in the momentum space is observed by varying the excitation energies.
\end{abstract}
\maketitle

\section{\label{sec:intro} INTRODUCTION}
Hand full of iron based superconductors have been exposed since the discovery of high $T_c$ superconductivity in LaFeO$_{1-x}$F$_x$As~\cite{Kamihara2008b}. But a heated debate on the mechanism of pairing glue is still running among the researchers. At an early stage of the discovery several theoretical reports undoubtedly argued that the mechanism of superconductivity in these materials is related to the short range spin-fluctuations (SFs) induced by the Fermi surface nesting between the disconnected Fermi sheets at the Brillouin zone center and the zone corner~\cite{Mazin2008a,Graser2010, Chubukov2008,Kuroki2009b}, this is further supported by several experimental observations~\cite{Brouet2009,Inosov2010,Analytis2010}.  Uncovering the nodal superconductivity in some class of ferropnictides questions whether the mechanism of superconductivity is based on the Fermi surface nesting~\cite{Yamashita2009c, Fletcher2009a, Dong2010a}. Nowadays, there is a growing consensus that the coupling of strong orbital fluctuations to the spin fluctuations may serve as the pairing mechanism in these superconductors~\cite{Saito2010,Stanescu2008,Kontani2010,Moreo2009a,Borisenko2012c}. 

The first non-magnetic Fe based superconductor LiFeAs, with a $T_c$ of 18 K does not exhibit a spin density wave order (SDW) and also shows an absence of nesting conditions,  challenging the existed theories of superconductivity in the Fe-based superconductors~\cite{Wang2008,Borisenko2010}. On the other hand, the NaFeAs superconductor belonging to the same 111-family,  shows a SDW order co-existing with superconductivity~\cite{Chu2009}. Interestingly, the issues on SC gap structure and gap sizes are still under debate for this compound. A recent London penetration depth data revealed a nodal gap structure in an optimally Co doped NaFeAs superconductor~\cite{Cho2012}, whereas the thermal conductivity measurements showed a fully gaped superconductivity~\cite{Zhou2012}. On the other hand, an ARPES study reported SC gap of 6.5 meV for the hole pockets and of 6.8 meV for the electron pockets~\cite{Liu2011}. Whereas the tunneling spectroscopy measurements suggested an average gap of 4.5 meV~\cite{Cai2012, Yang2012}. Nevertheless,  most of the reports suggested that the NaFe$_{1-x}$Co$_x$As superconductor is in close proximity to a strong-coupling superconductivity.

In this letter we report a systematic study on the electronic structure and SC gaps in the optimally doped  NaFe$_{0.95}$Co$_{0.05}$As superconductor.  Our results show a maximum gap $\Delta$ = 3.3 $\pm$ 0.3 meV on hole pockets and is of 2.9 $\pm$ 0.3 meV on electron pockets. From temperature dependent ARPES measurements we show that the SC gap closes above the transition temperature ($T_c$), suggesting an absence of pseudogaps in these compounds.  Our results clearly demonstrate that the gap evolution with temperature follow the BCS gap function near the $\Gamma$, $Z$, and $M$ high symmetry points. From photon energy dependent measurements we also observe an almost isotropic superconductivity along $k_z$ direction in the momentum space.

\section{\label{sec:exper}  EXPERIMENTAL DETAILS}
Angle-resolved photoemission spectroscopy (ARPES) is a vital tool to study the electronic structure and momentum and temperature dependent SC gaps. All of our measurements were carried out at the UE-112 beam-line equipped with 1$^3$ ARPES end station located in BESSY II (Helmholtz zentrum Berlin) synchrotron radiation center~\cite{Borisenko2012a, Borisenko2012b}. Photon energies for the measurements were varied between 15 eV to 50 eV. The energy resolution was set between 3 meV to 6 meV depending on the excitation energy. Data were recorded at a chamber vacuum of the order of 9 $\times$ 10$^{-11}$ mbar and the sample  temperature was varied between 1 K to 25 K. We employed various photon polarizations in order to extract the complete nature of the electronic structure. Single crystals of NaFe$_{0.95}$Co$_{0.05}$As were grown using the self-flux method and the samples showed superconductivity at a  transition temperature  $T_c$ = 18 K. 
  
\begin{figure}[t]
	\centering
		\includegraphics[width=0.45\textwidth]{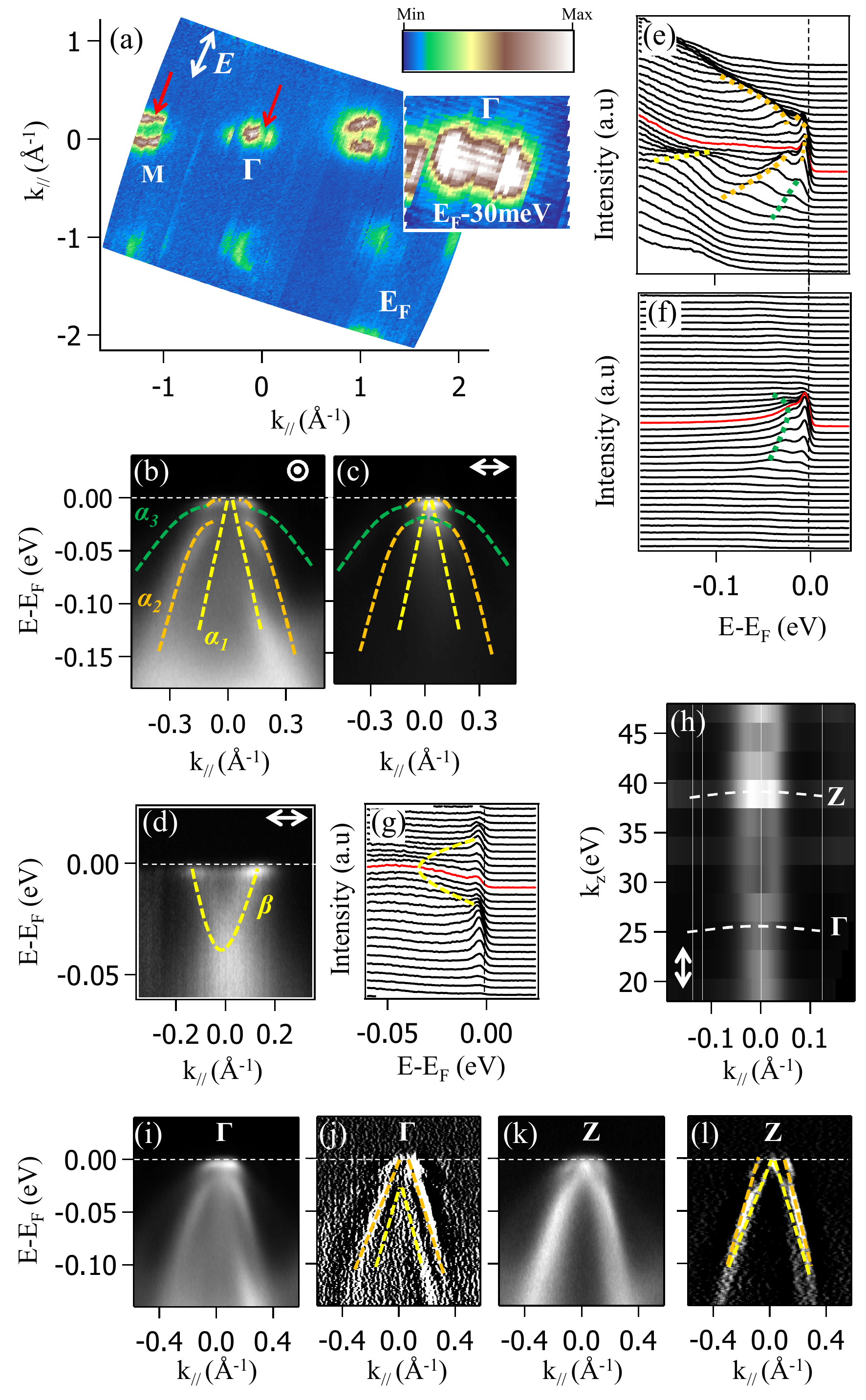}
	\caption{(color online) (a) is the FS map taken in the $k_x-k_y$ plane measured using $s$-polarized light with an excitation energy $h\nu$=80 eV, and inset is the map around $\Gamma$ taken at 30 meV below the Fermi level ($E_F$). (b) and (c) are the energy distribution map (EDM) cuts  taken at $\Gamma$ (along the direction shown on FS map) measured using $p$- and $s$-polarized lights, respectively. (d) is EDM cut taken at the $M$-point. Dashed lines on the EDMs are guides to the eye,  representing electronic band dispersions. (e), (f), and (g) are the energy distribution curves (EDCs) of EDMs shown in (b), (c), and (d), respectively. (h) is the FS map taken in the $k_{\parallel}-k_z$ plane measured using $p$-polarized light. (i) and (k) show the resultant EMDs from the addition of EDMs measured using $s$- and $p$-polarized light taken at $k_z$=0  and $\pi$ at the zone center, respectively. (j) and (l) are the respective second derivatives of the EDMs show in (i) and (k). In the figure $p$- and $s$-polarized photons are indicated by double circles and double sided arrow, respectively.}
		\label{1}
\end{figure}

\begin{figure}[t]
	\centering
		\includegraphics[width=0.45\textwidth]{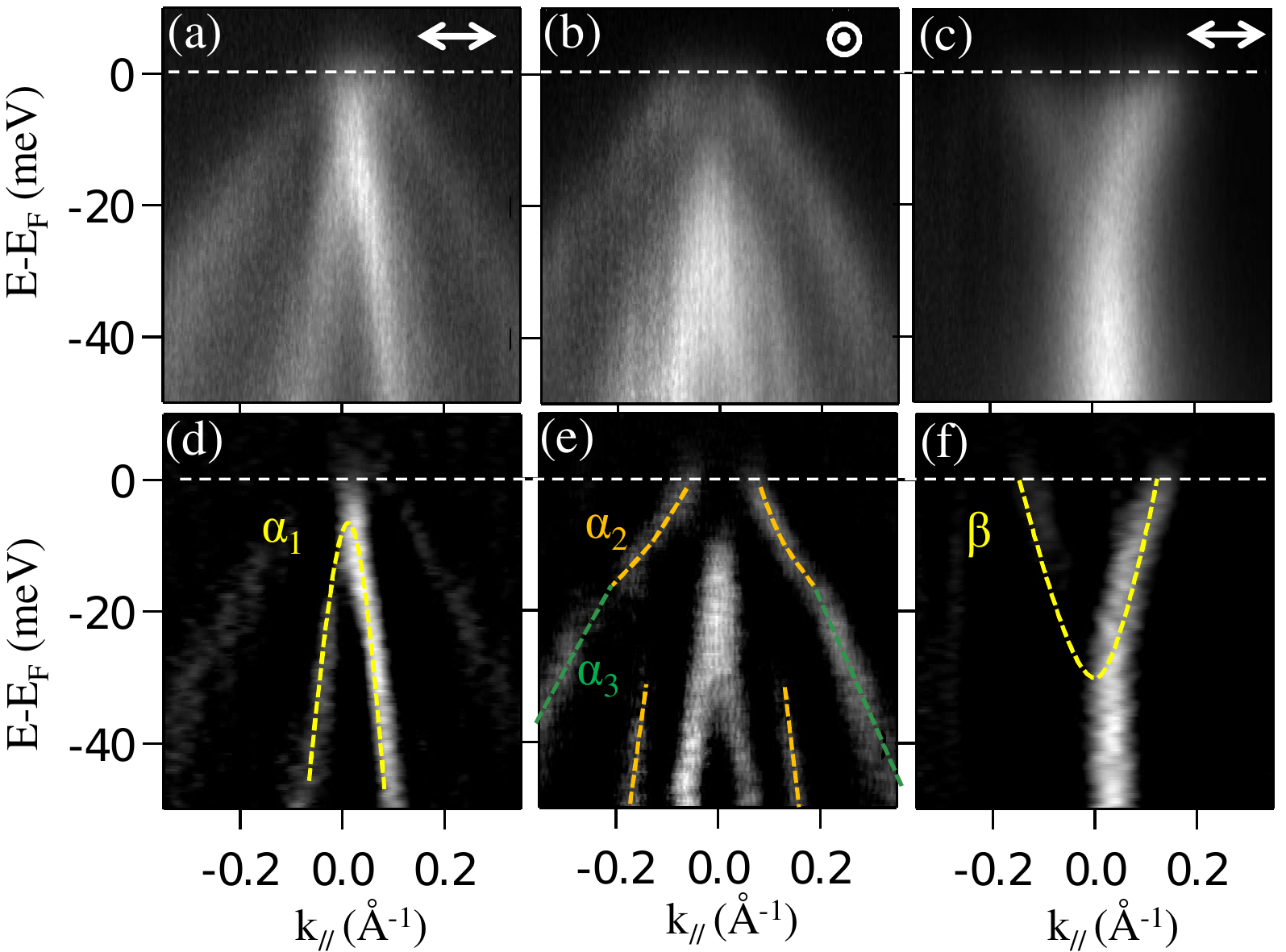}
	\caption{(color online) Top panel shows the EDMs measured with an excitation energy $h\nu$ = 20 eV when the sample is in  normal state ($T$ = 23 K). Bottom panel shows second derivatives of the respective EDMs.}
	\label{2}
	\end{figure}

\section{\label{sec:res}  RESULTS AND DISCUSSSION} 
Figure~\ref{1}(a) depicts the Fermi surface (FS) map measured using an excitation energy $h\nu$ = 80 eV and s-polarized photons in the $k_x-k_y$ plane. The data were recorded at a sample temperature of 1 K. We observe hole-like Fermi sheets at the zone center  and an electron-like Fermi sheet at the zone corner. Figures~\ref{1}(b) and (c) show energy distribution map (EDM) cuts taken at the zone center along the direction shown by red arrow at $\Gamma$ in Fig.~\ref{1}(a) which were measured using $p$- and $s$-polarized photons, respectively. From the EDM cuts we could extract three hole-like band dispersions $\alpha_1$, $\alpha_2$, and $\alpha_3$ at the zone center. Similarly, from an EDM cut taken at the zone corner [see Fig.~\ref{1}(d)]  we could observe an electron like band dispersion $\beta$. At the zone center we can see that the band $\alpha_3$ hybridizes with $\alpha_2$ and [see Figs.~\ref{1}(b),~\ref{1}(c), and~\ref{2}(e)], leads to a gap opening (E$_g$) for the band $\alpha_2$.  We further show that the band $\alpha_3$ does not hybridize with $\alpha_1$ [see Figs.~\ref{2}(a) and ~\ref{2}(d)], and thus no gap is opened for this band. At the moment the physical origin of selective band hybridization is unclear for us, but a  plausible reason could be related to the orbital ordering of the hole pockets~\cite{Graser2008}. Figures~\ref{1}(e), (f), and (g) show energy distribution curves (EDCs) of the EDMs that are shown in (b), (c) and (d),  respectively. Applying measuring geometry and photon polarization dependent selection rules we could assign ${xz/yz}$ orbital character for the bands $\alpha_1$, $\alpha_2$, and $\beta$ and  for the band $\alpha_3$ we assign ${xy}$ orbital character. 

Further, Fig.~\ref{1}(h) depicts the Fermi surface map taken at the zone center in the $k_{\parallel}-k_z$ plane measured using $p$-polarized light. The high symmetry points $\Gamma$ and $Z$ are identified using the formula $k_z=0.5123\sqrt{E_{kin}cos^2 (\theta)+V_0}$, here the inner-potential $V_0$ = 15 eV and the distance $\Gamma-Z=2\pi/c= 0.89 $\AA$^{-1}$, is calculated using the lattice parameter $c$ = 7.09 \AA ~\cite{Chu2009}. From the EDM cut taken at $\Gamma$-point [Fig.~\ref{1}(i) and (j)] we can see that the band $\alpha_2$ disperses up to the Fermi level ($E_F$) but posses a negligible Fermi vector, and $\alpha_1$ stays well below $E_F$. On the other hand, at Z [Fig.~\ref{1}(k) and (l)] we could observe the band $\alpha_2$ crosses $E_F$ at a Fermi vector $k_F$ = 0.082 $\pm$ 0.01 \AA$^{-1}$ and $\alpha_1$ disperses up to $E_F$ and touches it. These results suggest a substantial $k_z$ dispersion for the bands $\alpha_1$ and $\alpha_2$ at the zone center. 

Hole pockets at the zone center and electron pockets at the zone corner resemble the common electronic structure of all iron pnictides. However, a gap opening at the zone center makes a complicated electronic structure near the Fermi level in these compounds when compared to the other Fe-based superconductors~\cite{Fink2009}. Nevertheless, our results on the electronic structure are qualitatively in agreement with the band structure calculations and as well as with the published ARPES studies on the NaFeAs related compounds~\cite{Yi2011,Zhang2012a,Deng2009a,Kusakabe2009}.

\begin{figure}[ht]
	\centering
		\includegraphics [width=0.45\textwidth] {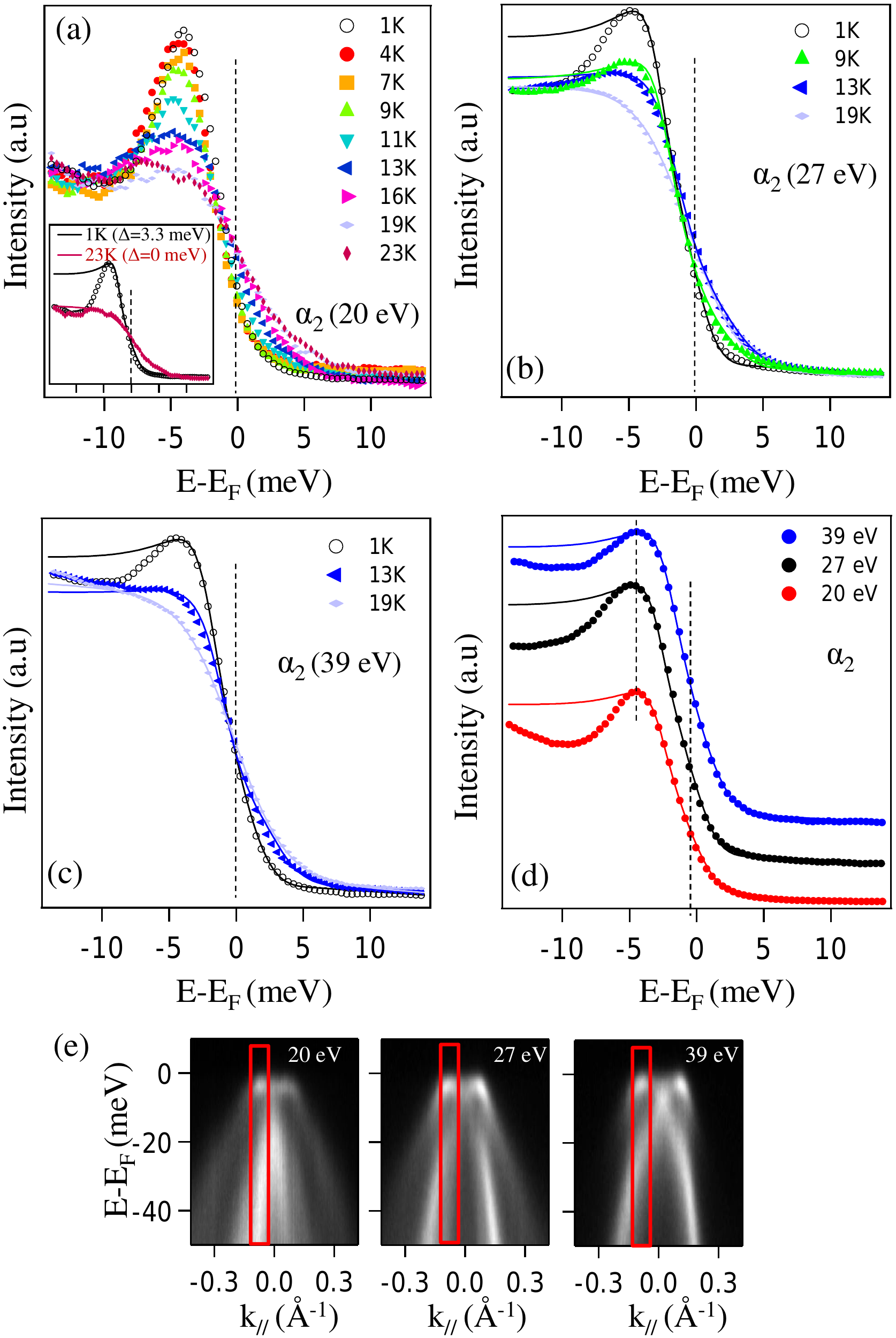}
	\caption{(color online) Integrated energy distribution curves (IEDCs) of the band $\alpha_2$ as a function of temperatures measured with photon energy 20 eV (a), 27 eV (b),  and 39 eV (c). (d) shows the IEDCs as a function of excitation energy measured at a sample temperature of 1 K. Solid curves in figure represent Dynes function fitting to the IEDCs (see text). (e) shows the ranges of momentum integration (red blocks) used to extract the EDCs shown in (a), (b) and (c) }
	\label{3}
	\end{figure}

\begin{figure}[htbp]
	\centering
		\includegraphics [width=0.45\textwidth] {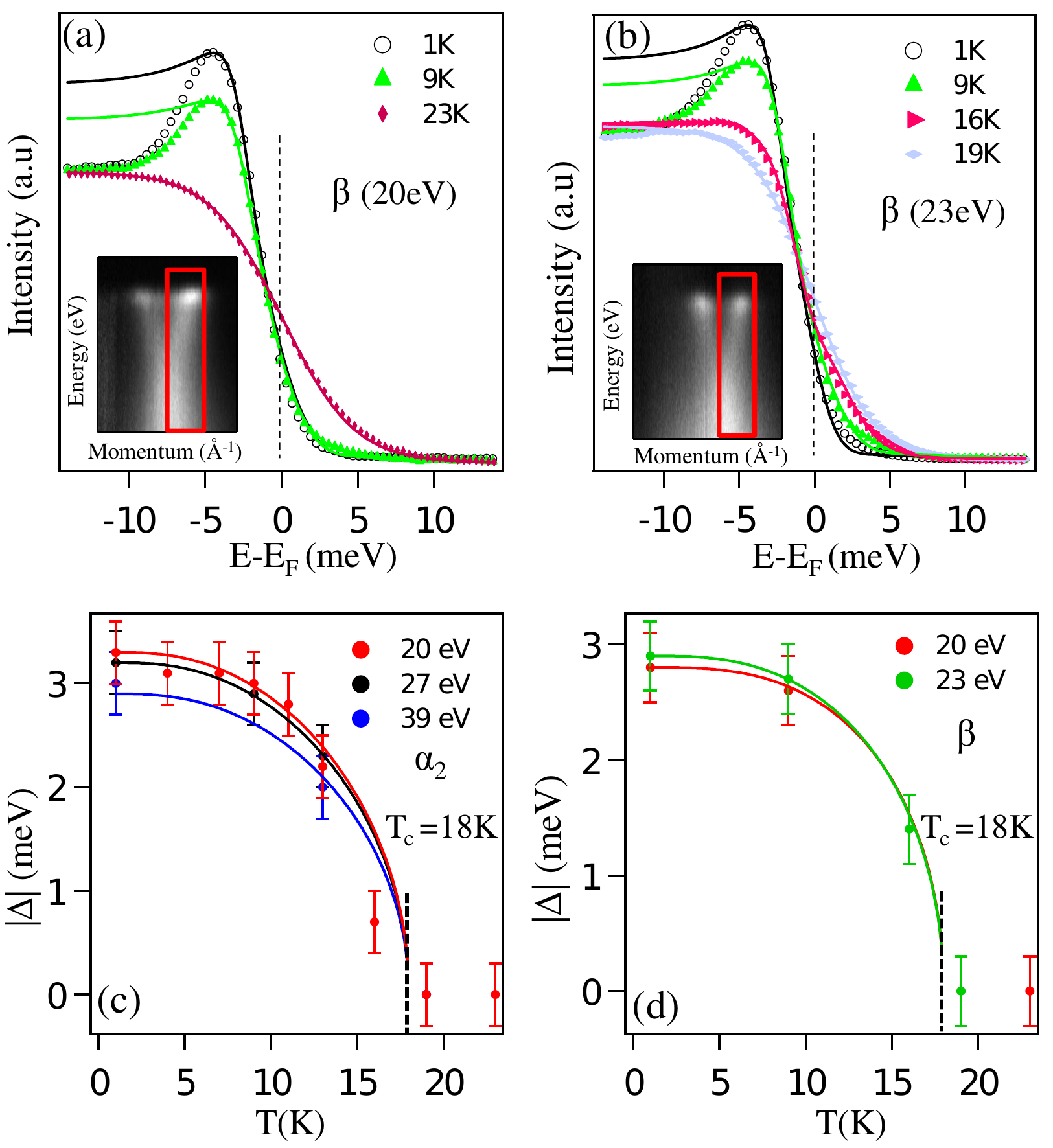}
	\caption{(color online) IEDCs are taken as a function temperature for the band $\beta$ measured with photon energy 20 eV (a) and 23 eV (b). Insets in (a) and (b) show the range of momentum integration (red block) used to extract the EDCs. (c) is SC gap evolution with temperature measured at various photon energies for the band $\alpha_2$. (d) is the same as (c) but taken from the band $\beta$.}
	\label{4}
\end{figure}

In order to extract the SC gap evolution with temperature we have performed temperature dependent ARPES measurements at the zone center and zone corner. In Figs.~\ref{3}(a), (b) and (c) we show integrated energy distribution curves (IEDCs) of the band  $\alpha_2$ as a function of temperature measured using the excitation energies h$\nu$ = 20, 27, and 39 eV [see Fig.~\ref{1}(h) for corresponding $k_z$ vectors in the momentum space], respectively. From Figs.~\ref{3}(a)-(c) one can clearly see that the coherent states near the Fermi level  gradually decrease up on increasing the sample temperature during a crossover from SC phase to a normal state, and no coherent states are observed above the critical temperature $T_c$ = 18 K. The gaps were extracted by fitting IEDCs with the Dynes function (solid curves). Dynes function fit to an IEDC is robust and complete analysis can be found elsewhere~\cite{Evtushinsky2009}. We obtained a superconducting gap  $\Delta$ = 3.3 $\pm$ 0.3 meV at a sample temperature of 1 K using the photon energy 20 eV. Applying the  similar procedure we extracted the gaps at various temperature and plotted them in Fig.~\ref{4}(c). Similarly, the gap evolutions with temperature for other photon energies are also plotted in Fig.~\ref{4}(c). Using the temperature dependent  BCS gap function,  $\Delta (T)= \Delta (0) \tanh[ \frac{\pi}{\delta_{sc}} \sqrt{a (\frac{\Delta C}{C}) (\frac{T_C}{T}-1)}]$,  we could nicely fit all the plots as shown in Fig.~\ref{4}(c).  Here, $\delta_{sc}=\frac{\Delta}{K_B T_c}$. And the constant $a (\frac{\Delta C}{C})$ is a specific heat jump in the SC phase, which is taken to be $\approx 1$ under the weak-coupling limit~\cite{Gross1986, Inosov2010a}. 

Figures~\ref{4}(a) and (b) show IEDCs of the band $\beta$ measured using the photon energies 20 eV and 23 eV, respectively at various temperatures. Applying the above mentioned fitting procedure we obtained a SC gap of 2.9$\pm$ 0.3 from the 20 eV data and 2.8$\pm$ 0.3 meV from the 23 eV data when the sample is at 1 K. The gap evolution with temperature is plotted in Fig.~\ref{4}(d) for both photon energies. Most importantly, all the bands show gap closing above $T_c$, which is in well agreement with Ref.~\onlinecite{Zhou2012a} where they reported absence of pseudogaps in the optimally Co doped NaFeAs superconductor and further demonstrated that the pseudogaps exist only in the overdoped compounds. Furthermore, our results support the in-plane momentum independent superconductivity in these compounds with a gap ratio $\Delta_{\alpha_2}/\Delta_{\beta}$ = 1.17 (calculated from the 20 eV data), and is consistent with the previous reports as well~\cite{Zhou2012a,Liu2011}.


\begin{figure}[ht]
	\centering
		\includegraphics [width=0.50\textwidth] {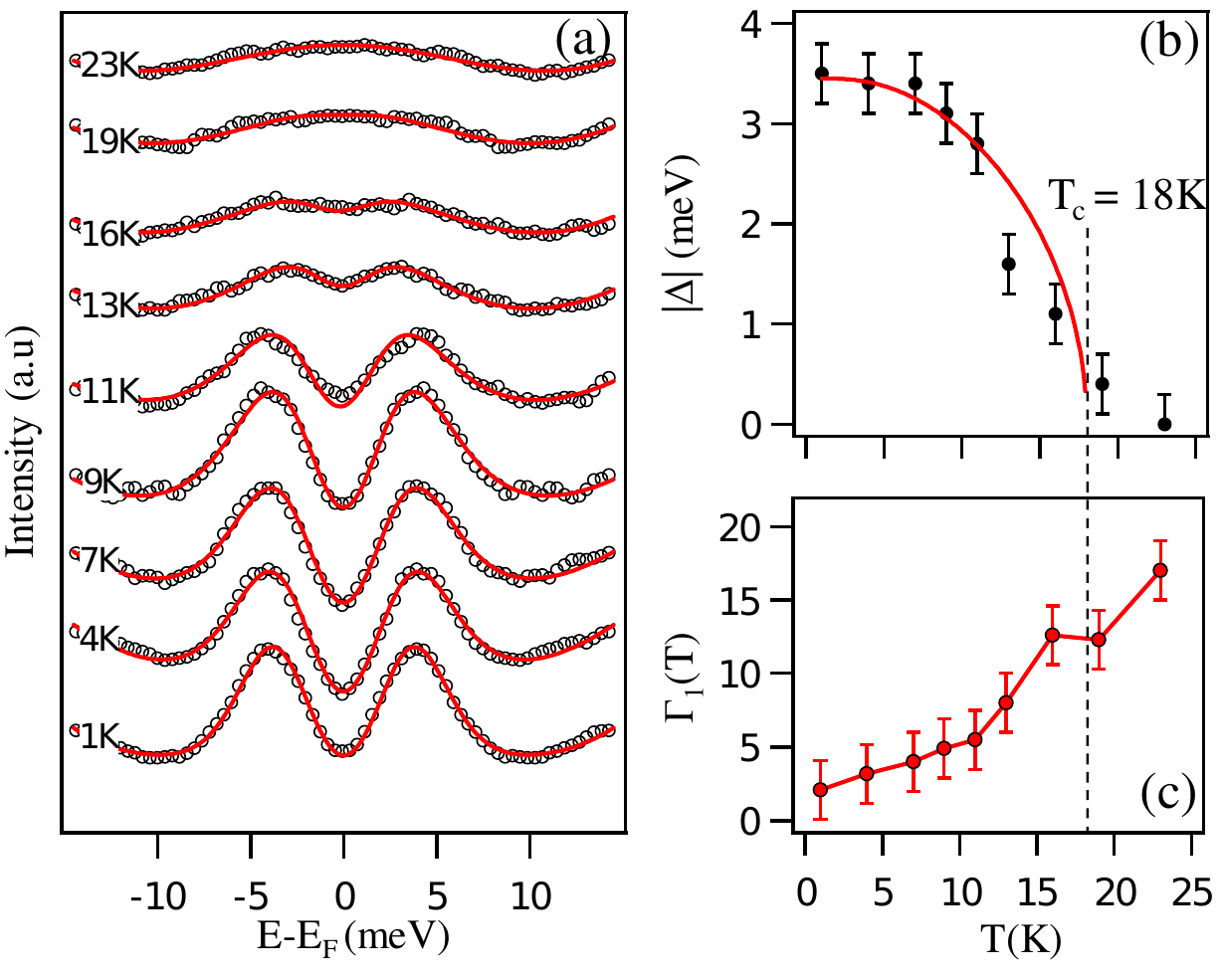}
	\caption{(color online) (a) Symmetrized IEDCs of the band $\alpha_2$ measured with the photon energy 20 eV. (b) SC gap evolution with temperature exctracetd from (a). (c) Single-particle scattering rates as a function of temperature. }
	\label{5}
\end{figure}

The main and important observation from our data is that  the SC gap sizes at both the zone center and corner are significantly smaller than the sizes (6.5 meV for the hole pockets and 6.8 meV for the electron pockets) reported from a similar ARPES technique~\cite{Liu2011}. In order to understand discrepancies in the gap sizes, we have followed the same procedure that is used in Ref.~\onlinecite{Liu2011}. In Fig.~\ref{5}(a) we show symmetrized IEDCs of the band $\alpha_2$ measured with  photon energy 20 eV at various temperatures. The symmetrized IEDCs are fitted using a spectral function of the self-energy, $\Sigma (k,\omega) = -i\Gamma_1+\Delta^2/[(\omega+i0^+)+\epsilon(k)]$~\cite{Norman1998}. Here $\Gamma_1$ is the single-particle scattering rate and is considered $\omega$ independent. The derived SC gaps in this method are again well in agreement with the ones presented in Fig.~\ref{4}(c),  and plotted them against sample temperature as shown Fig.~\ref{5}(b). Next, in Fig.~\ref{5}(c) we show the scattering rates as a function of temperature. One can clearly see in Fig.~\ref{5}(c) that the scattering rates are strongly temperature dependent. Furthermore, a sharp decrease below $T_c$ showing the quality of the crystal. We found $\Gamma_1$ = 2.2 meV for the holes and of 4.5 meV for the electrons (data is not shown) at a sample temperature of 1 K. These are much smaller than the scattering rates, 7 meV for the holes and 9 meV for the electrons reported in Ref.~\onlinecite{Liu2011}. Therefore from these observations we suggest that the larger gaps in Ref.~\onlinecite{Liu2011} may not entirely linked to superconductivity due to their existence well above $T_c$ with large scattering rates.  Most of the reports on SC gaps in NaFe$_{1-x}$Co$_{x}$As estimated from the coherence peak position~\cite{Liu2011, Cai2012, Yang2012}. This method generally overestimates the gap size due to the convoluted spectral resolution function~\cite{Evtushinsky2012}.  Recent tunneling spectroscopy measurements on the optimally doped NaFe$_{1-x}$Co$_{x}$As reported a gap of $\approx$ 4.5 meV, calculated from the distance between two coherence peaks~\cite{Cai2012, Yang2012}. On the other hand, we also observe the coherence peak around 4.3 meV for all the bands reported in this article. However, various fitting methods revealed the gap smaller than the coherence peak position.


	
Next we will discuss the $k_z$ dependent SC gaps at the zone center. In  Fig.~\ref{3}(d)  we show IEDCs measured using  the photon energies 20, 27, and 39 eV. We obtained a gap of 3.2$\pm$0.3 meV with 27 eV and 3$\pm$0.3 meV with 39 eV. From Fig.~\ref{1}(h) we can notice that the photon 27 eV is close to the $\Gamma$-point and 39 eV is close to the $Z$-point in the momentum space. Therefore, our results are suggesting a $k_z$ independent superconductivity in these compounds with a gap ratio $\Delta_{\Gamma}/\Delta_{Z}$ = 1.07. A recent ARPES study on Tl$_{0.63}$K$_{0.37}$Fe$_{1.78}$Se$_2$ also showed isotropic SC gaps along the $k_z$ direction~\cite{Wang2012}. It has been reported that in iron pnictides the SC gap at $Z$ is smaller compared to the size at $\Gamma$~\cite{Zhang2010g,Evtushinsky2012}. This is explained as the inter-orbital pairing interaction between the hole and electron pockets is relatively weaker compared to the intra-orbital pairing interaction ~\cite{Saito2010,Graser2008}. A transformation of the orbital character from $xz/yz$ to $z^2$ has been observed while going from $\Gamma$ to $Z$ in the case of 122 systems~\cite{Thirupathaiah2010}. Therefore in $k_z$ = $\pi$ plane we have an inter-orbital interaction between hole ($z^2$) and electron pockets ($xz/yz$)  and in the $k_z$= 0 plane an intra-orbital interaction is realized between hole ($xz/yz$) and electron pockets ($xz/yz$). Hence anisotropic SC gaps are expected in the $k_z$ direction. Whereas our results show an isotropic superconductivity in these compounds at the zone center, and we do not expect a different behavior at the zone corner. 

It has been suggested that the Fermi surface nesting is crucial for acquiring superconductivity in iron pnictides~\cite{Kemper2010, Graser2008}. On the other hand, several ARPES studies show imperfect nesting in the doped iron pnictides~\cite{Thirupathaiah2011, Thirupathaiah2010, Liu2011a}.  In the present study we do see an absence of nesting conditions between $\alpha_2$ and $\beta$ in the $k_z$=0 plane. Whereas there could  be a substantial nesting in the $k_z$= $\pi$ plane. But in both planes we observe equivalent SC gaps at the zone center. Now this questions whether the FS nesting plays an important role in iron pnictides. Mostly, it appears that the superconductivity is proximity to the Fermi surface topological transition, which seems to be a common feature in all the iron based conductors~\cite{Borisenko2010,Borisenko2012}. Nevertheless, the presented $k_z$ independent SC gaps in this article can not be explained by a simple FS nesting mechanism. Perhaps a theory of orbital-fluctuations coupled to the spin-fluctuations might be needed in order to understand the in-plane and out-of-plane momentum independent superconductivity in these compounds~\cite{Saito2010,Stanescu2008,Kontani2010,Moreo2009a}.



\section{\label{sec:con}  CONCLUSIONS}

In conclusion, the electronic structure of NaFe$_{0.95}$Co$_{0.05}$As superconductor is different from other ferropnictides in the sense that a gap ($E_g$) is opened at the zone center due to the hybridization between the hole-like bands. Our observations on the electronic structure are in qualitative agreement with the DFT band structure calculations and published ARPES reports on these compounds. Obtained SC gaps at various high symmetry points are within the range of 2.8-3.3 meV, and corresponding $\frac{\Delta}{K_B T_c}$ are in the range of 1.8-2.1. This is clearly suggesting that the optimally electron doped NaFe$_{0.95}$Co$_{0.05}$As superconductor is close to the BCS weak-coupling limit (1.76).  Gap closing above the transition temperature ($T_c$) shows the absence of pseudogaps. Gap evolution with temperature follow the BCS gap equation near the $\Gamma$, $Z$, and $M$ high symmetry points.  Furthermore, we observed in-plane and out-of-plane momentum independent superconductivity in the NaFe$_{0.95}$Co$_{0.05}$As superconductor. Thus, our observations put this compound in the weak-coupling limit of the 111-family of iron pnictides, such as LiFeAs superconductors.  

\section*{ACKNOWLEDGMENTS} 
Work at IFW was supported by the DFG (Grant BE1749/13, SPP1458). S.W acknowledges funding by DFG through the Emmy-Noether program (Grant WU595/3-1). S.W., I.M., and M.R. acknowledge support from ERA.Net RUS (project acronym: FeSuCo, ID: STProjects-245 ). I.M. and M.R. acknowledge funding by RFBR (project No: 12-03-01143-a).

\bibliographystyle{apsrev}
\bibliography{NaFeCoAs-PRB}
\end{document}